\documentclass[letterpaper, 10 pt, conference]{ieeeconf}  % Comment this line out
\IEEEoverridecommandlockouts                              % This command is only
\title{\LARGE \bf
Mesh Processing Strategies and Fractals for Three Dimensional Morphological Analysis of a Granitic Terrain using IRS LISS IV and Carto DEM 
}

\author{K. Seshadri$^{1}$ and M. Naresh Kumar$^{2}$% <-this % stops a space
\thanks{*ESRI User Conference, Hyderabad September 2018. This work is supported by National Remote Sensing Centre (ISRO)}% <-this % stops a space
\thanks{$^{2}$M. Naresh Kumar is with Geoportals and WebGIS services Group, National Remote Sensing Centre, India
        {\tt\small nareshm at ieee.org}}%
\thanks{$^{1}$K. Seshadri is with the Geo-sciences group, Remote Sensing Applications Area, 
        National Remote Sensing Centre, India
        {\tt\small seshadri\_k at ieee.org}}%
}

\usepackage{graphicx}
\begin{document}

\maketitle
\thispagestyle{empty}
\pagestyle{empty}

%%%%%%%%%%%%%%%%%%%%%%%%%%%%%%%%%%%%%%%%%%%%%%%%%%%%%%%%%%%%%%%%%%%%%%%%%%%%%%%%
\begin{abstract}

Virtual Reality (VR) enabled applications are becoming very important to visualize the terrain features in 3D. In general 3D dataset generated from high resolution satellites and DEM occupy large volumes of data. However, light weight datasets are required to create better user experiences on VR platforms. So, the present study develops a methodology to generate datasets compatible to VR using Indian Remote Sensing satellite (IRS) sensors. A Linear Imaging Self-Scanning System - IV (LISS IV) with 5.8 m spatial resolution and Carto DEM are used for generating the 3D view using Arc environment and then converted into virtual reality modeling language (VRML) format. In order to reduce the volume of VRML dataset a quadratic edge collapse decimation method is applied which reduces the number of faces in the mesh while preserving the boundary and/or normal. A granitic terrain in the south-west part of Hyderabad comprising of dyke intrusion is considered for generation of 3D VR dataset, as it has high elevation differences thus rendering it most suitable for the present study. Further, the enhanced geomorphological features such as hills and valleys, geological structures such as fractures, intrusive (dykes) are studied and found suitable for better interpretation.
\end{abstract}
%%%%%%%%%%%%%%%%%%%%%%%%%%%%%%%%%%%%%%%%%%%%%%%%%%%%%%%%%%%%%%%%%%%%%%%%%%%%%%%%
\section{INTRODUCTION}
The three-dimensional (3D) datasets generated from high resolution remote sensing sensors and digital elevation model reveal earth’s surface and expose its interiors thus making it extremely useful in understanding earth system processes \cite{c1}. Virtual Reality (VR) Platforms such as Google’s Daydream, Oculus Rift, Sony PlayStation VR, Samsung Gear VR offer a very good platform for realistic visualization. The Google Earth VR (https://vr.google.com/earth/) provides new experiences of urban environment in 3D. The VR applications have been extensively used in predicting traffic patterns \cite{c2}, Urban planning \cite{c3}, understanding earth quakes \cite{c4}, granitic terrains \cite{c5}.  Recent developments in virtual reality hardware and software has resulted in shifting of focus from PC based systems to light weight headgears providing a better human computer interface. However, there is a strong need to generate light weight 3D datasets which offer better user experiences at reduced computational costs. 

The elevation or z value plays a crucial role in generating 3D views. The elevation values are in general derived from a stereo pair using a stereo strip triangulation method and stored in digital elevation model (DEM) format. To get a colored 3D image the texture information derived from a multi spectral sensor is draped over the DEM. The  DEM used in the present study is generated from a 500 km stereo pairs of Cartosat-1 which is available at 10m interval spacing and high resolution dataset of Resoursat-2 LISS IV sensor is used to get tone and texture variations. Esri ArcScene 10.5 environment is used to combine the DEM with the LISS IV data to obtain the 3D view.

3D datasets are stored as Mesh consisting of vertices, faces and texture. They are stored as VRML, X3D, ply, Wavefront Obj etc formats. Using ArcScene VRML format is generated for the present dataset to process further.

The size of a mesh is directly proportional to sensor resolution of the base dataset. Thus, the high resolution datasets have  larger mesh sizes due to more dense points (vertices and faces). Hence, mesh simplification is required to remove redundancies retaining the original quality.  

Presently, 3D model of a granitic terrain covering south of Ibrahimpatnam, Telangana is generated from IRS LISS IV and Carto DEM. Further, a light weight version is created using  simplification procedures given in http://www.meshlab.net

\section{Methods}
Esri ArcScene environment provides set of tools for generation and visualization of 3D \cite{c6}. Also, it provides interfaces to navigate and interact with 3D features and raster data. ArcScene provides relatively faster navigation of complex 3D features through in memory operations and OpenGL API. Initially, the Resourcesat-2 LISS IV data and Carto DEM are imported into ArcScene. Later, 3D view is generated by setting base height of LISS IV image to DEM   using floating on custom surface option. Further, VRML format is generated through export option in Arcscene.

The generated VRML is imported into Meshlab environment for simplification by applying a quadratic edge collapse decimator approach. The parameters used in the simplification process include target numbers of faces, quality threshold, planar simplification, preserving normal and boundaries of the mesh. They help in constraining the simplification process, retaining shape of the triangles and mesh boundaries. The 3D dataset is then converted to Wavefront obj format for visualization in VR gadgets and applications. 

\begin{figure}[thpb]
      \centering
            \includegraphics[width=\linewidth]{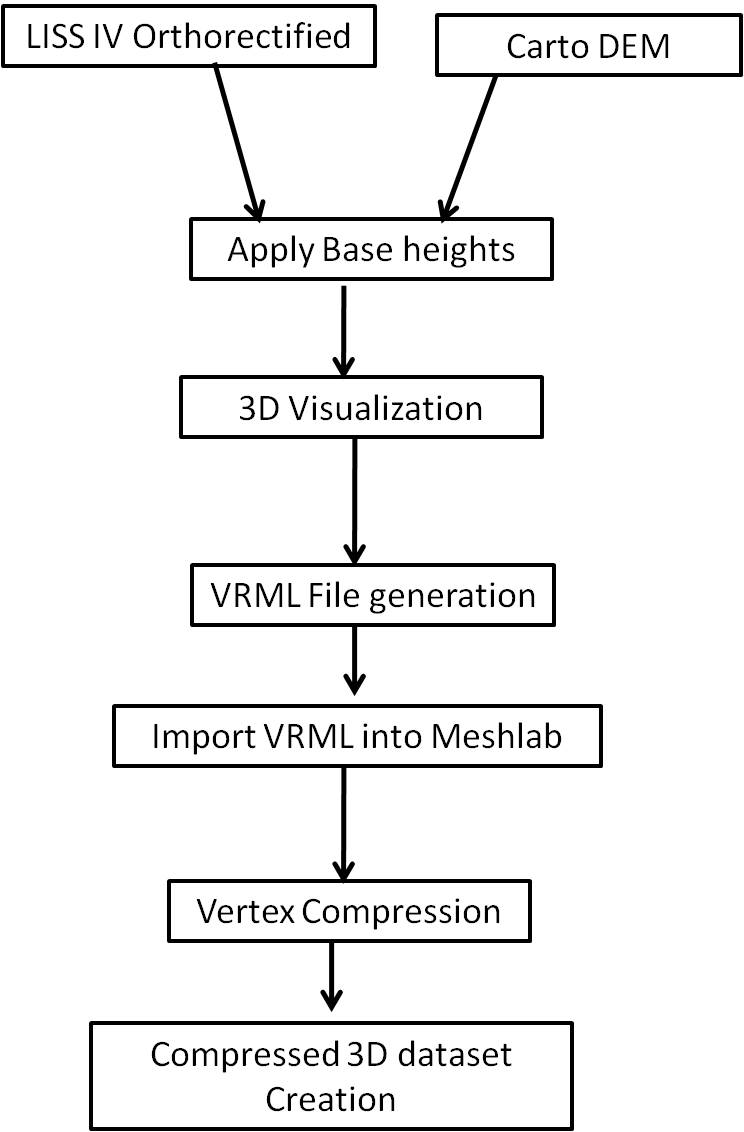}
      \caption{Proposed Methodology}
      \label{fig1}
   \end{figure}

\section{Results}
\begin{figure}[thpb]
      \centering
            \includegraphics[width=\linewidth]{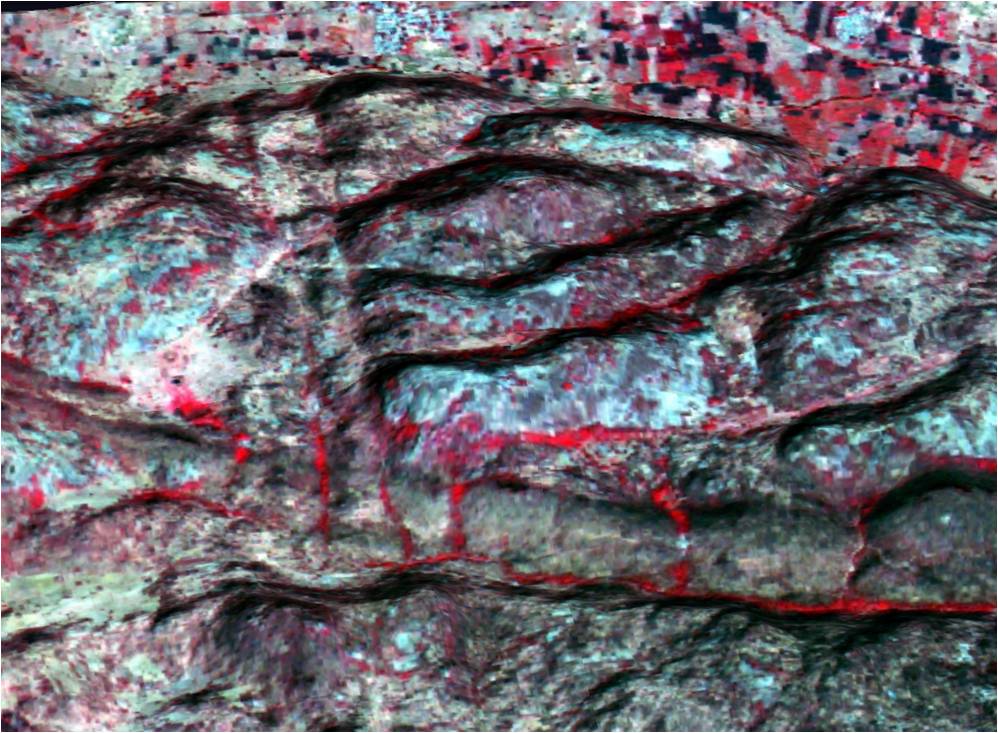} 
            \includegraphics[width=\linewidth]{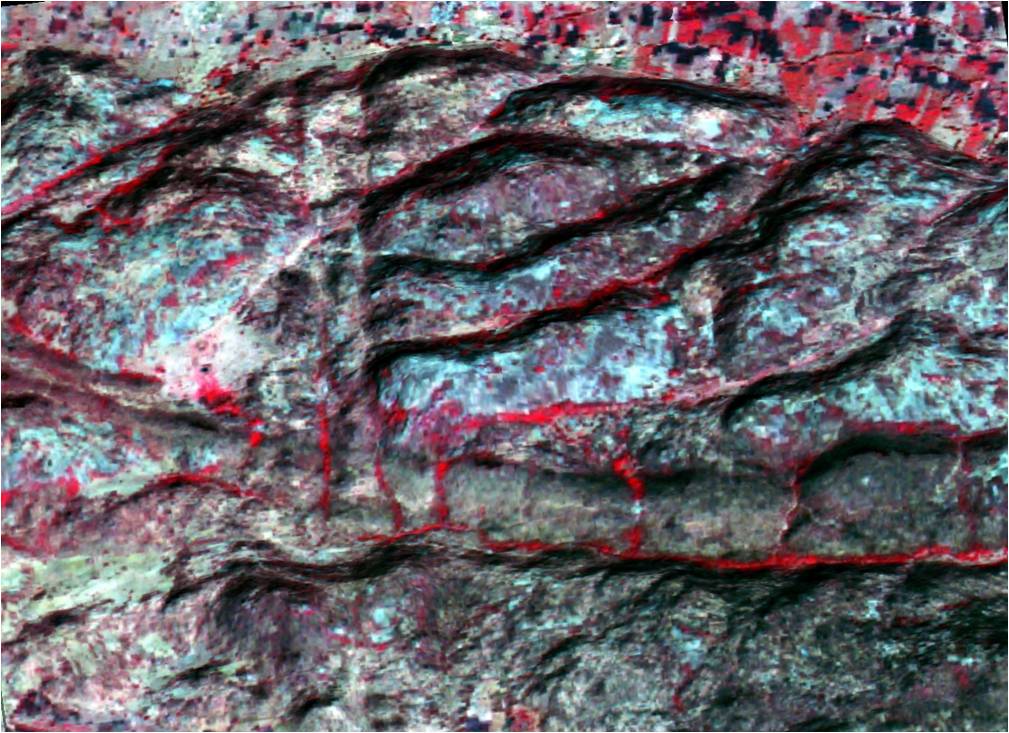}
      \caption{Granitic terrain covering south of Ibrahimpatnam, (a) before and (b) after compression}
      \label{fig2}
   \end{figure}

The 3D image of LISS IV covering 20 sq km is converted to VRML using Carto DEM in ArcScene which occupies 9 MB of disk space having 322202 vertices and 632468 faces. After vertex compression the file size is reduced to 3 MB with 18042 vertices and 30116 faces. The original and compressed images are shown in Figure \ref{fig2}. Around 5\% reduction is seen in the number of vertices and faces before and after compression. Also, a 50\% reduction in file sizes is observed between VRML and obj formats. The results are tabulated in Table \ref{tab1}.

\begin{table}[h]
\caption{Number of faces and vertices before and after compression  
}
\label{tab1}
\begin{center}
\begin{tabular}{|c||c||c||c|}
\hline
Dataset & Sizes (KB) & Vertices & Faces\\
\hline
Original & 9331 & 322202 & 632468 \\
Compressed &3497 & 18042 & 30116 \\
\hline
\end{tabular}
\end{center}
\end{table}

\section{CONCLUSIONS}
The ArcScene environment is useful for generating 3D perspective views. However, due to lack of further processing techniques generation of compressed formats suitable for VR applications requires additional tools such as Meshlab. The mesh simplification has resulted in fewer number of vertices and faces, thereby reducing the file sizes by 50\%. It is found that the 3D perspective views in VR apps has made it possible to better interpret the granitic terrain.

\section*{ACKNOWLEDGMENT}
The authors profusely thank Director, NRSC for his support. Thanks to Deputy Director, RSA and DPPA\& WAA, NRSC for their encouragement and support.

\end{document}